\documentclass[%
 reprint,
 groupedaddress,
superscriptaddress,
 amsmath,amssymb,
 aps,
]{revtex4-1}

\usepackage{mathrsfs}
\usepackage{textcomp}
\usepackage{units}
\usepackage{array,mathtools}
\usepackage{stmaryrd}
\usepackage{scrextend}
\usepackage{nicefrac}
\usepackage{multirow}
\newcolumntype{C}{>{$}c<{$}}
\AtBeginDocument{
\heavyrulewidth=.08em
\lightrulewidth=.05em
\cmidrulewidth=.03em
\belowrulesep=.65ex
\belowbottomsep=0pt
\aboverulesep=.4ex
\cmidrulesep=\doublerulesep
\cmidrulekern=.5em
\defaultaddspace=.5em
}
\usepackage{booktabs}
\usepackage[caption=false]{subfig}
\usepackage{graphicx}% Include figure files
\usepackage{dcolumn}% Align table columns on decimal point
\usepackage{bm}
\usepackage{xspace}

\newcommand{\strue}{$S_{(a)}$\xspace} 
\newcommand{\smeta}{$S_{(b)}$\xspace}
\newcommand{\ttrue}{$T_{(a)}$\xspace}
\newcommand{\tmeta}{$T_{(b)}$\xspace}
\newcommand{\stmeta}{$(S/T)_{(b)}$\xspace}
\newcommand{\sttrue}{$(S/T)_{(a)}$\xspace}

\newcommand{\smetaY}{$S_{(c)}$\xspace}
\newcommand{\tmetaY}{$T_{(c)}$\xspace}
\newcommand{\stmetaY}{$(S/T)_{(c)}$\xspace}
\usepackage{xcolor}
\usepackage[normalem]{ulem}

\newcommand{\editor}[2]{%
  \expandafter\newcommand\csname #1note\endcsname[1]{%
    \textcolor{#2}{(\textbf{#1:} ##1)}}%
  \expandafter\newcommand\csname #1\endcsname[1]{%
    \textcolor{#2}{##1}}%
  \expandafter\newcommand\csname #1cancel\endcsname[1]{%
    \textcolor{#2}{\sout{##1}}}%
  \expandafter\newcommand\csname #1change\endcsname[2]{%
    \textcolor{#2}{\sout{##1} ##2}}%
  \newenvironment{#1text}{\color{#2}}{\color{black}}
}

\editor{LB}{violet}
\editor{NM}{orange}
\editor{MK}{red}
\editor{IT}{green}

\begin{document}

\title{Hybridization driving distortions and multiferroicity in rare-earth nickelates}

\author{Luca Binci}
\affiliation{Theory and Simulation of Materials (THEOS), and National Centre for Computational Design and Discovery of Novel Materials (MARVEL), École Polytechnique Fédérale de Lausanne, CH-1015 Lausanne, Switzerland}
\author{Michele Kotiuga}
\affiliation{Theory and Simulation of Materials (THEOS), and National Centre for Computational Design and Discovery of Novel Materials (MARVEL), École Polytechnique Fédérale de Lausanne, CH-1015 Lausanne, Switzerland}
\author{Iurii Timrov}
\affiliation{Theory and Simulation of Materials (THEOS), and National Centre for Computational Design and Discovery of Novel Materials (MARVEL), École Polytechnique Fédérale de Lausanne, CH-1015 Lausanne, Switzerland}
\author{Nicola Marzari}
\affiliation{Theory and Simulation of Materials (THEOS), and National Centre for Computational Design and Discovery of Novel Materials (MARVEL), École Polytechnique Fédérale de Lausanne, CH-1015 Lausanne, Switzerland}
\affiliation{Laboratory for Materials Simulations, Paul Scherrer Institut, 5232 Villigen PSI, Switzerland}
\date{\today}

\begin{abstract}
For decades transition-metal oxides have generated a huge interest due to the multitude of physical phenomena they exhibit. In this class of materials, the rare-earth nickelates, $R$NiO$_3$, stand out for their rich phase diagram stemming from complex couplings between the lattice, electronic and magnetic degrees of freedom. Here, we present a first-principles study of the low-temperature phase for two members of the $R$NiO$_3$ series, with $R=$ Pr, Y. We employ density-functional theory with Hubbard corrections accounting not only for the on-site localizing interactions among the Ni--$3d$ electrons ($U$), but also the inter-site hybridization effects between the transition-metals and the ligands ($V$). All the \textit{U} and \textit{V} parameters are calculated from first-principles using density-functional perturbation theory, resulting in a fully \emph{ab initio} methodology. Our simulations show that the inclusion of the inter-site interaction parameters $V$ is necessary to simultaneously capture the features well-established by experimental characterizations of the low-temperature state: insulating character, antiferromagnetism and bond disproportionation. On the contrary, for some magnetic orderings the inclusion of on-site interaction parameters $U$ alone completely suppresses the breathing distortion occurring in the low-temperature phase and produces an erroneous electronic state with a vanishing band gap. In addition -- only when both the \textit{U} and \textit{V} are considered -- we predict a polar phase with a magnetization-dependent electric polarization,  supporting very recent experimental observations that suggest a possible occurrence of type-II multiferroicity for these materials.

\end{abstract}

\maketitle

%Example commands for comments:\\
%     Add something: \LB{addition}\\
%     Cancel: \LBcancel{cancel}\\
%     Cancel and add: \LBchange{change}{add}\\
%     Note: \LBnote{note}\\
%Different colors: MK, IT, LB (see \textbackslash editor)

\section{Introduction}
Rare-earth nickelates \emph{R}NiO$_3$ (\emph{R} = rare-earth element) are a class of materials exhibiting a particularly rich phase diagram~\cite{Medarde1997}. Except LaNiO$_3$, these materials undergo a metal-insulator phase transition (MIT) coinciding with a symmetry-lowering distortion of the crystal structure~\cite{Alonso1999, Alonso2000}. 
This distortion changes the space group from $Pbnm$, in which all Ni sites are crystallographically equivalent, to $P2_1/n$ -- which possesses two inequivalent Ni sites -- because of the activation of a breathing mode, that contracts and expands the NiO$_6$ octahedra in a rock-salt pattern. The MIT temperature, $T_\mathrm{MIT}$, is a monotonically decreasing function of the size of the rare-earth cation ($\text{Lu}\rightarrow\text{La}$), that coincides with the Nèel temperature $T_{\text{Nèel}}$ for $R$ = Nd, Pr, and vanishes for $R$ = La. In the last years rare-earth nickelates have attracted a great deal of attention, especially for their application within epitaxial heterostructures~\cite{Middey2016}, which have been shown to possess crossovers between different magnetic orders~\cite{Hepting2018}, the possibility to manipulate $T_\mathrm{MIT}$ through electric fields~\cite{Scherwitzl2014}, and exotic types of phases transitions~\cite{Post2018,Liu2013}. 
Recently, superconductivity has been induced via apical oxygen deintercalation together with Sr doping in $R$NiO$_3$ with $R$ = Nd, Pr~\cite{Li2019,Osada2020}. Finally, recent experiments have suggested a low-temperature multiferroic phase signaled by additional Raman-active modes compatible with the loss of the inversion center in the antiferromagnetic (AFM) phase~\cite{Ardizzone2021}.

\begin{figure*}[t]
\centering
\includegraphics[width=17cm]{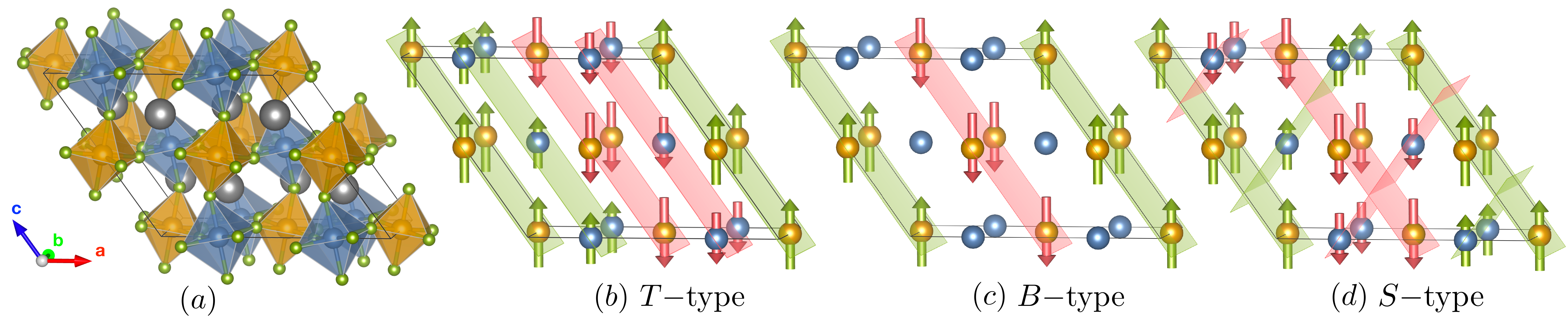}
\caption{(\textit{a}) Unit cell of $R$NiO$_3$ in the $P2_1/c$ space group doubled along the direction of the lattice vector $\mathbf{a}$ in order to accommodate the different AFM orderings studied here. Blue (orange) octahedra contain the short (long)-bond nickels, Ni$_\mathrm{S}$ (Ni$_\mathrm{L}$), respectively. The grey and the green atoms are the rare-earth elements ($R$ = Pr or Y) and the O atoms, respectively. The other panels show the three  AFM orderings investigated: (\textit{b})~$T$-type, (\textit{c})~$B$-type, and (\textit{d})~$S$-type; here, for clarity, only the Ni atoms are shown. The arrows represent the directions of the magnetic moments on the Ni atoms and their lengths are proportional to their modulus. In the $S/T$-type orderings all of the Ni atoms have a magnetic moment, while in the $B$-type ordering the nickels with shorter Ni$-$O bond lengths are nonmagnetic. The green and red planes are guides for the eyes to visualize planes of spin-up and spin-down moments, respectively. Rendered using \textsc{VESTA}~\cite{Momma2008}.}
\label{magn_exp}
\end{figure*} 

Besides the MIT, the nature of the low-temperature insulating phase has been the subject of intense study. An orbital ordering induced by a Jahn-Teller distortion was ruled out in favour of charge disproportionation~\cite{Mazin2007}; this picture had also reached experimental consensus~\cite{Alonso1999, Medarde2008}. However, such charge disproportionation has not manifested in first-principles calculations~\cite{Park2012}, which show a negligible difference in the $3d$ electronic occupations of the two crystallographically inequivalent Ni$_\mathrm{S}$ and Ni$_\mathrm{L}$ (defined below), both of which have approximately 2 electrons in the $e_g$ states, and therefore 8 electrons in the $3d$ manifold -- an effect clearly explained within the negative-feedback charge regulation mechanism discussed in Ref. \cite{Raebiger2008}. Hereafter we use the subscripts S/L, to indicate the Ni ion in the smaller/larger NiO$_6$ octahedrons, which are contracted/expanded by the breathing-mode distortion. As rare-earth nickelates are considered to be negative charge-transfer insulators~\cite{Zaanen1985}, it has been proposed that, starting from a configuration with one ligand hole ($\underline{L}$) on every nickel, the transition $(3d^8\underline{L})\,(3d^8\underline{L})\rightarrow (3d^8\underline{L}^2)_\mathrm{S}\,(3d^8)_\mathrm{L}$ takes place at the MIT~\cite{Johnston2014}. According to this picture, one NiO$_6$ octahedron contracts and the two $e_g$ electrons on Ni$_\mathrm{S}$ couple with the two oxygen holes, forming a low-spin state. The other NiO$_6$ octahedron expands and ends up in a high-spin state owing to the Hund's coupling. Similar conclusions have been drawn from an analysis based on density-functional theory (DFT) + dynamical mean field theory (DMFT)~\cite{Park2012, Haule2017} (that named the state a ``site-selective" Mott state), DFT+$U$ studies~\cite{Varignon2017, Hampel2017}, multi-band many-body Hamiltonians solved with Hartree-Fock~\cite{Mizokawa2000, Johnston2014} and exact diagonalization methods~\cite{Green2016}. 

Despite such intense theoretical effort, several open questions still remain. In particular, the precise magnetic order remains elusive. Indeed, there is wide experimental agreement only on the magnetic propagation vector $\mathbf{Q}$, which encodes the periodicity of the magnetic order with respect to the primitive cell of the space group. However, the precise values of the magnetic moments on the Ni atoms have not yet been unambiguously determined~\cite{Garcia-Munoz1994, Alonso1999}. From first-principles calculations it has been reported that only small values of Hubbard $U\lesssim 2$ eV successfully stabilize an AFM order that is both compatible with the experimental \textbf{Q} and energetically favourable, with respect to the ferromagnetic (FM) order~\cite{Varignon2017, Hampel2017, Mercy2017}. Still, the values for the Hubbard parameter were tuned by scanning different values, and were not calculated rigorously with a well-determined theoretical protocol, thus preventing predictions to be entirely nonempirical. More importantly, these calculated magnetic orders have systematically zero magnetic moments on Ni$_\mathrm{S}$~\cite{Badrtdinov2021}. As it is shown here, this feature is not compatible with a possible multiferroic nature for $R$NiO$_3$, which was first computationally predicted~\cite{Giovannetti2009} and more recently experimentally highlighted by Raman spectroscopy measurements, which revealed the emergence of additional phonon modes in the low-temperature phase, compliant with an inversion symmetry breaking required for a type II multiferroic phase~\cite{Ardizzone2021}.  In order to computationally demonstrate such multiferroic phase, it is necessary to establish a coupling between a nonvanishing electric polarization \textbf{P} and the magnetic degrees of freedom, i.e. to prove a dependence of \textbf{P} on the simulated magnetic order. A group theory analysis  \cite{PerezMato2016} has shown that a nonzero \textbf{P} can be obtained if the magnetic moments on the Ni$_\mathrm{S}$ atoms are different from zero, for instance within the $S$- and/or $T$-types magnetic orderings~\cite{Giovannetti2009} (see Fig.~\ref{magn_exp}). Unfortunately, for $U\simeq 2$ eV these latter magnetic patterns are stable only in a range of amplitudes for the breathing mode distortion that are much smaller than the experimental ones~\cite{Badrtdinov2021}. Larger values of $U$ may lead to converged $S/T$-types orderings, but they could also cause a FM instability~\cite{Varignon2017, Hampel2017}. 

In this paper we use a fully first-principles DFT+$U$+$V$ approach that includes both on-site ($U$) interactions to capture the electronic localization and inter-site ($V$) interactions to account for hybridization effects~\cite{Campo2010}. We calculate the Hubbard parameters self-consistently using density-functional perturbation theory (DFPT)~\cite{Timrov2018, Timrov2021}, thus resulting in a parameter-free theory. Our calculated Hubbard $U$ parameters range from 8 to 9~eV and are numerically different for the two crystallographically inequivalent $\mathrm{Ni}_\mathrm{S}/\mathrm{Ni}_\mathrm{L}$. Using these high values of $U$, after an iterative approach aimed to determine the self-consistent crystal structure and Hubbard parameters, we find that within DFT+$U$~\cite{Anisimov1991, Dudarev1998} -- having only on-site electronic contributions -- the investigated $R$NiO$_3$ structures would be found in a FM ground state, in agreement with other theoretical studies \cite{Varignon2017,Hampel2017} and in contrast with experimental observations. Moreover, the DFT+$U$ calculated AFM orders of the $S/T$-types are also found to have a vanishing breathing mode distortion and zero band gap, in disagreement with experiments. Conversely, by including the inter-site Hubbard $V$ interactions between the Ni and O sites, the results change considerably: in particular, we find qualitatively different final states for the $S/T$ orderings. 
In addition to the fact that some of them exhibit the expected AFM instability against FM ordering, with DFT+$U$+$V$ we always recover an insulating character, consistently with experiments. This feature allows us to evaluate the electric polarization \textbf{P}, which is found to be magnetization-dependent, thus providing additional support for the emergence of multiferroicity \cite{Giovannetti2009,Ardizzone2021}. We also simulate the magnetic order with zero magnetic moments on Ni$_\mathrm{S}$ (i.e. the $B$-type ordering, see Fig.~\ref{magn_exp}) and show that this is not able to generate a polar instability, thus keeping the materials in a non-polar phase with a vanishing \textbf{P}. Our results for PrNiO$_3$ and YNiO$_3$ -- sitting on opposite ends of the $R$NiO$_3$ phase diagram -- are qualitatively similar, suggesting that these numerical findings could also be extended to other members of the rare-earth nickelate series.
%\LBchange{We therefore argue that these numerical evidences for Pr and Y nickelates -- sitting on opposite ends of the $R$NiO$_3$ phase diagram -- support the possibility of extending our results to other members of the rare-earth nickelate series.}{}

\begin{table*}[]
\centering
\caption{\label{table1} Structural and electronic properties of PrNiO$_3$ and YNiO$_3$ in the low-temperature phase for different magnetic orderings computed with different DFT+Hubbard schemes and as measured in experiments. Positive $\Delta E$ means that FM is lower in energy than AFM, and vice versa. Three types of the AFM order are considered: $T$, $S$, and $B$ types (see Fig.~\ref{magn_exp}). The $V$ interaction parameters are different within a given Ni$-$O octahedron because of the different Ni$-$O bond lengths yielding three inequivalent sites (O$_1$, O$_2$ and O$_3$). However, the differences between the different values of $V$ within a given octahedron are very small and for conciseness we report their average values; The full results are available in the Materials Cloud Archive~\cite{MaterialsCloudArchive2022}. The metallic (M) or insulating (I) character for each case is also indicated.}
\bgroup
\def\arraystretch{1.2}
\setlength\tabcolsep{0.1in}
\begin{tabular}{lccccccc}
\toprule
\toprule
Method&AFM&$U_\mathrm{S}/U_\mathrm{L}$ (eV)&$\langle V_\mathrm{S}/V_\mathrm{L}\rangle_\mathrm{avg}$ (eV)&$m_\mathrm{S}/m_\mathrm{L}$ ($\mu_\mathrm{B}$)&$\langle\ell_{\mathrm{S}}/\ell_{\mathrm{L}}\rangle_\mathrm{avg}$ (\AA)&$\Delta E$ (meV/f.u.)& Met/Ins\\
\midrule
\multicolumn{8}{c}{PrNiO$_3$}\\
%\cline{2-7}
%DFT&$S$&&&0.00/0.00&1.92/1.92&\\
%&$T$&&&0.00/0.00&1.92/1.92&\\
%&$B$&&&0.00/0.00&1.92/1.92&\\
\midrule
DFT+$U$&$S$&9.40/9.40&&1.54/1.53&2.00/2.00&156&M\\
&$T$&9.32/9.33&&1.52/1.53&1.99/2.00&137&M\\
&$B$&9.19/8.19&&0.00/1.68&1.88/2.03&200&I\\
%\midrule
DFT+$U$+$V$&\strue&8.76/9.61&0.76/0.90&0.64/1.65&1.89/2.03&$-26$&I\\
&\ttrue&8.77/9.60&0.76/0.90&0.69/1.68&1.90/2.03&$-23$&I\\
&\smeta&9.78/9.33&1.02/0.95&1.14/1.61&1.94/2.02&71&I\\
&\tmeta&9.78/9.42&1.01/0.96&1.15/1.61&1.94/2.02&75&I\\
&$B$&9.60/8.49&0.84/0.71&0.00/1.67&1.88/2.03&90&I\\
%\midrule
Expt. \cite{Gawryluk2019}&&&&&1.92/1.97& & I\\
\midrule
\multicolumn{8}{c}{YNiO$_3$}\\
\midrule
%DFT&$S$&&&0.00/0.78&1.92/1.95&\\
%&$T$&&&0.00/0.75&1.92/1.95&\\
%&$B$&&&0.00/0.78&1.92/1.95&\\
%\midrule
DFT+$U$&$S$&9.38/9.38&&1.51/1.51&2.00/2.00&115&M\\
&$T$&9.21/9.20&&1.49/1.49&2.00/2.00&72&M\\
&$B$&9.17/8.06&&0.00/1.68&1.88/2.04&143&I\\
%\midrule
DFT+$U$+$V$&\strue&8.38/9.49&0.65/0.80&0.15/1.70&1.88/2.05&$-94$&I\\
&\ttrue&8.40/9.50&0.66/0.81&0.18/1.70&1.88/2.05&$-91 $&I\\
&\smetaY&9.31/9.25&0.82/0.79&1.32/1.32&1.99/1.99&7&I\\
&\tmetaY&9.35/9.30&0.84/0.82&1.33/1.33&1.99/1.99&6&I \\
&$B$&9.54/8.35&0.80/0.68&0.00/1.67&1.88/2.04&61&I\\
Expt. \cite{Alonso2001}&&&&&1.92/2.01& & I\\
\bottomrule
\bottomrule
\end{tabular}
\egroup
\end{table*}

\section{Results}
The AFM magnetic orderings studied in this paper are shown in Fig.~\ref{magn_exp}. All of these require a supercell of the 20-atom unit cell of the monoclinic $P2_1/n$ structure, which in turn is a $\sqrt{2}\times\sqrt{2}\times2$ supercell of the 5-atom cubic perovskite structure. %$Pm\bar{3}m$. 
Previously these orderings have been investigated using a 80-atom 2$\times$1$\times$2 supercell of the $P2_1/n$ space group, in order to reproduce the experimentally observed AFM propagation vector $\mathbf{Q}_{P2_1/n}=(\nicefrac{1}{2},0,\nicefrac{1}{2})$ \cite{Hampel2017,Varignon2017}. 
To reduce the computational cost\LB{s}, we have used the $P2_1/c$ setting of the space group 14, which has the same $\mathbf{a}$ and $\mathbf{b}$ primitive lattice vectors  as $P2_1/n$, but $\mathbf{c}_{P2_1/c}=\mathbf{a}+\mathbf{c}_{P2_1/n}$~\cite{Ardizzone2021}. In this setting the magnetic propagation vector becomes $\mathbf{Q}_{P2_1/c}=(\nicefrac{1}{2},0,0)$, only requiring a doubling of the 20-atoms $P2_1/c$ unit cell (instead of quadrupling it, as for the 20-atoms $P2_1/n$). 

In Table \ref{table1}, we report for each of the AFM configurations the results of the calculated magnetic moments $m$, total energy difference with respect to the FM configuration $\Delta E$, the average bond lengths for the Ni$-$O octahedra $\langle \ell\rangle_\mathrm{avg}$, and the computed values of the Hubbard interaction parameters $U$ and $V$. 
As shown in Table \ref{table1}, DFT+$U$+$V$ yields two states for each of the two $S/T$ types of orderings. Both of them are energetically degenerate in pairs (within numerical accuracy): the lowest-energy ones, that we denote as \sttrue, and the higher-energy (metastable) states named \stmeta and \stmetaY. As described in the Methods section these states are found with an iterative self-consistent loop alternating calculations of Hubbard parameters and crystal structure optimizations until convergence of the two is reached simultaneously. In general we find stable points within this workflow, except for the \sttrue states. For these latter, at each iteration the protocol produces a periodic reversal of the breathing mode amplitude, to which equivalent switches of the Hubbard parameters and magnetic moments are associated. Since the numerical values of $U$, $V$, Ni--O octahedral volumes and magnetic moments associated to Ni$_\mathrm{S}$/Ni$_\mathrm{L}$ are numerically always the same, together with the fact that the total energy does not exhibit any variation but a smooth convergence with the iterations, we argue that this does not create physical ambiguities, as all the main features of the final result are well-defined (for more information, see Supplementary Information~\cite{SupplementaryInformation}). This fact, however, underlines the need of a formulation for Hubbard-corrected DFT functionals able to take into account the variation of the Hubbard parameters with respect to atomic positions and strains, inasmuch as this dependence can be quite significant \cite{Kulik2011}.

%\LBchange{For the \sttrue states, the iterative protocol reaches a stalemate: from the large (small) NiO$_6$ octahedra, the smaller (larger) $U$ and $V$ are calculated, which in turn drive the contraction (expansion) of the corresponding NiO$_6$ octahedron in the subsequent structural relaxation, and so on [see Fig.~S2~(a) in the Supplementary Information (SI)~\cite{SupplementaryInformation}]. However, although the breathing mode oscillates at each iteration, it always keeps the same absolute amplitude. The total energies of the two crystal structures with the inverted breathing modes are almost equal (within numerical accuracy) as well as both the values of the magnetic moments and the Hubbard parameters corresponding to the small (large) octahedra. We  therefore concluded that these electronic states are in fact specular and reflect the same unique electronic ground state. The driving feature leading to this oscillation is the use of \emph{different} (i.e. site-dependent) Hubbard parameters for $\mathrm{Ni}_\mathrm{S}/\mathrm{Ni}_\mathrm{L}$, which, besides being physically expected from symmetry considerations, is correctly captured by our site-resolved linear-response scheme.}{}
\begin{figure*}[]
\centering
\includegraphics[width=17.5cm]{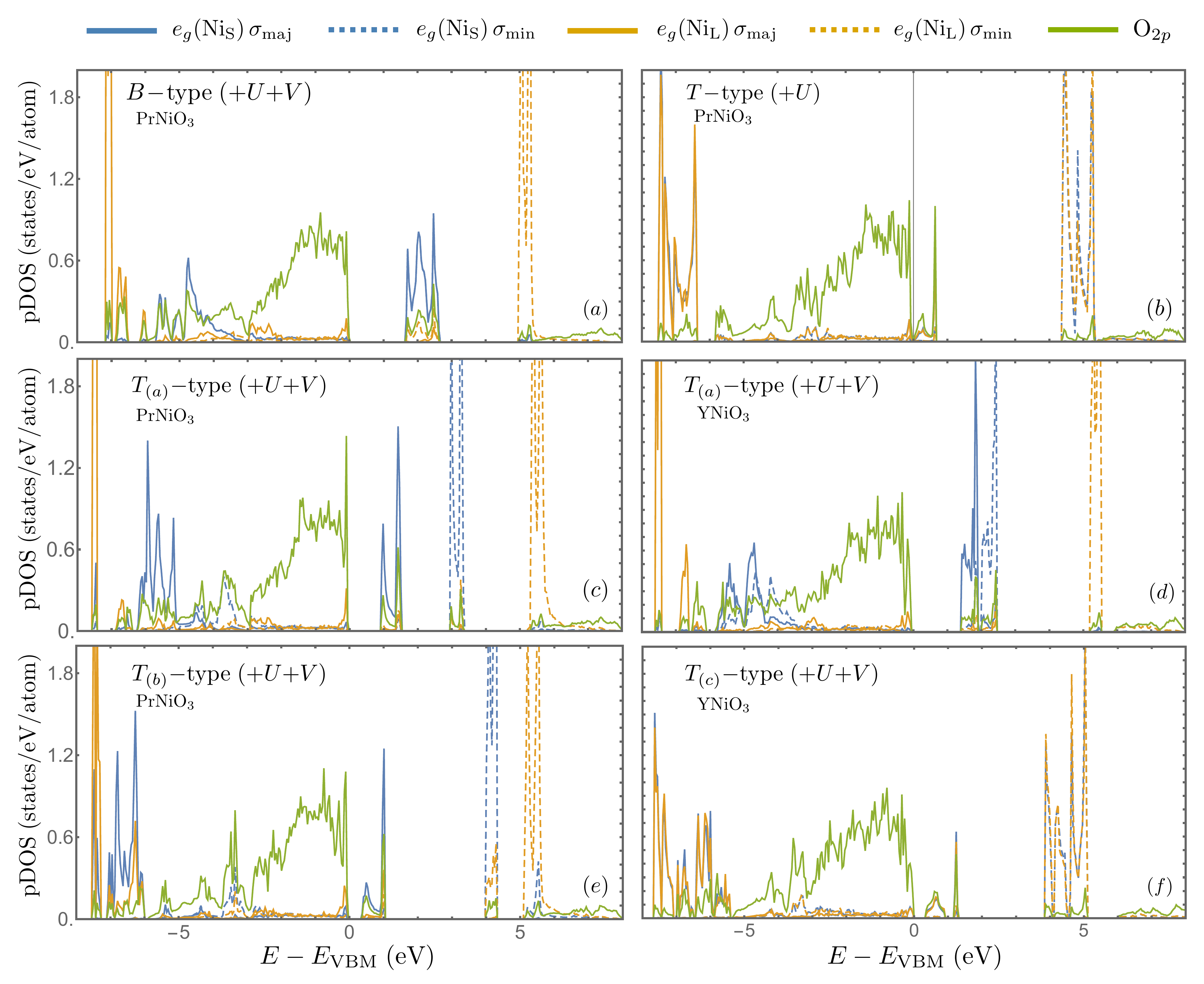}
\caption{pDOS of the relaxed crystal structure for each magnetic ordering. (\textit{a}), (\textit{b}):~\textit{B}-type and $T$-type, respectively, for PrNiO$_3$. (\textit{c}), (\textit{d}):~\ttrue-type and (\textit{e}), (\textit{f}):~\tmeta-type and \tmetaY-type, for both PrNiO$_3$ and YNiO$_3$. The Hubbard functionals employed to produce (\textit{a}), (\textit{c})--(\textit{f}) include both the $U$ and the $V$ interaction parameters, while in (\textit{b}) only the on-site $U$ is present. In the plots the Ni--$e_g$ states and the full O--$2p$ projection are shown. We averaged over inequivalent O atoms and over the two Ni($e_g$) states for each spin channel. Given the symmetry between spin-up/spin-down channels for AFM orderings, the notation for majority/minority (maj/min) spin channels is: for a Ni with positive magnetic moments ($m>0$ $\mu_\mathrm{B}$), $\sigma_\mathrm{maj}=\,\uparrow$ and $\sigma_\mathrm{min}=\,\downarrow$; vice versa for Ni with $m<0$ $\mu_\mathrm{B}$. The zero of energy corresponds to the valence band maximum.}
\label{pDOS}
\end{figure*} 

%It is found that the larger parameters $\{U_\mathrm{lrg},V_\mathrm{lrg}\}$ are extracted from $\mathrm{Ni}_\mathrm{S}$. Within the structural relaxation process, $\{U_\mathrm{lrg},V_\mathrm{lrg}\}$ cause an expansion of the associated NiO$_6$ octahedron and enhance the Ni local magnetic moment. The opposite effect occurs for $\{U_\mathrm{sml},V_\mathrm{sml}\}$. This mechanism further points out the large entanglement between the electron and lattice degrees of freedom that is present for these materials. On the contrary, the structurally converged \stmeta states have very similar $\{U_\mathrm{lrg},V_\mathrm{lrg}\}$ and $\{U_\mathrm{sml},V_\mathrm{sml}\}$ parameters and they are not sufficient to create the imbalance necessary to reach the same structurally NiO$_6$ disproportionated state. 

The total energies reported in the last column of Table \ref{table1} are calculated at the relaxed crystal structure with the corresponding self-consistent, site-dependent $U$ and $V$ Hubbard parameters, both for the AFM and the FM configurations. Since the final $U$ and $V$ are not numerically the same, it follows that the reference energies are different. This could in principle create ambiguities when comparing total energies calculated with functionals containing numerically different parameters. However, we regard the Hubbard term as a correction to the self-interaction error of the DFT base functional, selectively applied to the Ni--$3d$ (and O--$2p$) manifold of the corresponding ground state. According to this point of view it is natural that different electronic ground states will be characterized by different self-interaction errors (i.e. different curvatures with respect to the $3d$ electron occupancy, see Ref.~\cite{Cococcioni2005}). As a consequence we follow Ref.~\cite{Cococcioni2019} and argue that, for each of electronic state, the corresponding $U$ and $V$, aimed to remove such errors, must necessarily be different. 

We highlight that we use orthogonalized atomic states taken from the pseudopotential as Hubbard projectors (see Methods), while the calculations with $U\simeq 2$~eV by other works mentioned above in the text were carried out with orbital projectors derived from the projector augmented-wave (PAW) formalism~\cite{Bengone2000}. In general same values for the Hubbard parameters can have a different impact on the electronic structure when applied to different Hubbard projectors~\cite{Wang2016}. Therefore, to establish the consistency of  two computational setups in the same range of $U$ parameters, we have numerically verified for PrNiO$_3$ that with the empirical $U\simeq 2$~eV we stabilize a $B$-type AFM order -- even starting from the $S/T$-type orderings -- that is lower in energy than the FM configuration, in compliance with previous studies carried out with PAW projectors. 

In Fig.~\ref{pDOS} we show the projected density of states (pDOS) of the \ttrue and \tmeta types of orderings with and without the inter-site $V$, and the $B$-type ordering computed using DFT+$U$+$V$. Equivalent results for the $S$-type ordering are almost identical to the $T$-type ordering. Furthermore, the results for the $B$-type ordering obtained with and without inter-site $V$ present no qualitative differences and, thus for sake of conciseness we report only the DFT+$U$+$V$ results. 
%The results of the other calculations can be found in Sec.~S1 in the SI~\cite{SupplementaryInformation}.
In particular we plot the Ni--$e_g$ and the O--$2p$ states. The Ni $t_{2g}$ bands, corresponding to localized, fully-occupied states, yield narrow peaks located below $-5$~eV with respect to the highest occupied states and for clarity are not reported in the same figure (they can be found in SI~\cite{SupplementaryInformation}).
%We note that as the Ni$-$O bond lengths within a given octahedron are different, the point symmetry is lower than $O_h$ and, therefore, the $e_g$ and $t_{2g}$ symmetry labels are not strictly valid. However, since the deviation from the $O_h$ point group is small we use these higher-symmetry labels. The eigenvalues of the occupation matrix of the localised $3d$ manifold~\cite{Cococcioni2005} (for each spin channel) yield three fully occupied levels and two with fractional occupancies (see Supplementary Information), we identify the former with $t_{2g}$ states within the local-axis reference frame (rotated because of the octahedral tilting), and the latter with the corresponding $e_g$ states. We have thus projected these latter electronic states within such rotated $3d$ manifolds, as described in Ref.~\cite{Mahajan2021}, and they are reported in Fig.~\ref{pDOS} for PrNiO$_3$ and YNiO$_3$. 

In Table \ref{table2} we show the results of the calculated electric polarization \textbf{P} using the Berry phase approach \cite{King-Smith1993,Umari2002}. Because of the polar instability, during the structural optimization the crystal structure lowers its symmetry, and thus the space group changes. The newly obtained space groups are reported in the same table. For the $S/T$ orderings, we mention that we obtained a nonzero \textbf{P} also at the experimental crystal structure, in agreement with Ref. \cite{Giovannetti2009}, as the inversion symmetry-breaking is induced at the level of the {magnetic} space group.
%\ITnote{This lonely sentence looks awkward. Either we expand the discussion (e.g. providing more details) or remove it at all. I think we need to try the first option.}

%Finally in Table \ref{table2} the results of the calculated electric polarization \textbf{P} are shown, together with the lower-symmetry space groups to which the crystal structure relaxes upon ionic optimization. 

\section{Discussion}
\subsection{Electronic, crystal and magnetic structures}
\paragraph{DFT+$U$}
The $B$-type AFM ordering presents no relevant differences when examined with the two methods, i.e. with or without the inter-site parameter $V$: they show a band gap of $\simeq 1.4$ eV, very similar magnetic moments of $|m|\simeq1.7$ $\mu_\mathrm{B}$ on Ni$_\mathrm{L}$ and it is always found to be higher in energy than the FM configuration. As shown in Table~\ref{table1}, the breathing mode of the $S$- and $T$-type orderings  obtained with on-site interactions only (DFT+$U$) disappears, as the Ni$-$O bond lengths of the two structurally inequivalent  NiO$_6$ octahedra become equal. In addition, as shown in Fig.~\ref{pDOS}, the electronic states associated with these orderings display a semimetallic behaviour, with the $3d$ states of the two Ni atoms becoming equivalent. This state also exhibits almost the same values of $U$ interaction parameters and magnetic moments for the two inequivalent Ni sites. For this reason we conclude that within this \emph{ab initio} model, at the self-consistent crystal structure and Hubbard parameters, purely on-site interactions can not sustain the breathing-mode-based structural distortion. 

\paragraph{DFT+$U$+$V$}
Upon the inclusion of the inter-site Hubbard interactions between the nickels and the oxygens, the materials recover the experimentally observed insulating nature. The lowest-energy, semi-degenerate states \sttrue share similar features for PrNiO$_3$ and YNiO$_3$: they display significantly different magnetic moments and Ni$-$O bond lengths for the two inequivalent Ni sites. This difference in $m_\mathrm{S}/m_\mathrm{L}$ and $\langle\ell_\mathrm{S}/\ell_\mathrm{L}\rangle$ is larger for YNiO$_3$ than for PrNiO$_3$ mirroring the larger disproportionation observed for $R$NiO$_3$ with a smaller rare-earth ionic radius. Importantly for these states, the numerically larger (smaller) Hubbard parameters are applied to the larger (smaller) Ni--O octahedra. The opposite applies to the $B$ and \stmeta--types orderings. This difference turned out to be essential for stabilize \sttrue states as lower in energy than the FM state.  
%In addition, the pDOS of the \sttrue types shows an insulating behavior with a band gap $\simeq 1$~eV for PrNiO$_3$ and $\simeq 1.4$~eV for YNiO$_3$. Interestingly, for PrNiO$_3$, 1~eV is also the best fit parameter for the charge-transfer energy $\Delta=E(3d^{n+1}\underline{L})-E(3d^n)$ which realizes in Ref.~\cite{Mizokawa1995} an almost perfect match between a configuration-interaction calculation of a NiO$_6$ cluster model with adjustable parameters -- $\Delta$, $U$ and transfer integrals expressed as Slater-Koster parameters -- and the experimental X-ray absorption spectrum (XAS) in the same paper. From the same study, the extracted on-site Hubbard energy is also $U\simeq 7$ eV, that is closer to our calculated $U$ values than to those $U\lesssim 2$~eV empirically chosen with the aim of stabilizing the AFM order as the lowest energetic state. 
The metastable \stmeta for PrNiO$_3$ states exhibit a substantially smaller $\mathrm{Ni}_\mathrm{S}/\mathrm{Ni}_\mathrm{L}$ inequivalence because of a significant reduction of the breathing mode. They are different from the corresponding \stmetaY states obtained for YNiO$_3$, as in these latter the breathing mode amplitude shrinks to zero (see Table~\ref{table1}). In addition, both \stmetaY display different electronic occupations in the $e_g$ bands (see Supplementary Information) and different polar distortions with respect to \sttrue states, exemplified as opposite signs of some components of \textbf{P} between \smetaY and \strue (see Table \ref{table2}). 

The pDOS in Fig. \ref{pDOS} shows that the materials are negative charge-transfer insulators, since -- according to  definition -- the $2p$ electronic states of the oxygens stand between the upper/lower Hubbard bands of the $3d$ states of the two nickels. The negative character of charge-transfer is due to the presence of residual O$-{2p}$ states in the lowest conduction bands (the so-called hole-doped ligands)~\cite{Bisogni2016}. In all the panels in Fig.~\ref{pDOS}, the $e_g$ states of Ni$_\mathrm{L}$ [$e_g(\mathrm{Ni}_\mathrm{L})$] present concentrated spectral weights 5~eV above (minority-spin channel, $\sigma_\mathrm{min}$) and 7~eV below (majority-spin channel, $\sigma_\mathrm{maj}$) the valence-band maximum (VBM). The situation is similar for $e_g(\mathrm{Ni}_\mathrm{S})$ in DFT+$U$, as the two nickel sites become structurally equivalent [see Fig.~\ref{pDOS}~(b)]. A reduction in the magnetization on Ni$_\mathrm{S}$ 
produces a gradual downshift of the $e_g(\mathrm{Ni}_\mathrm{S})$ $\sigma_\mathrm{min}$ states, until they finally merge with the $e_g(\mathrm{Ni}_\mathrm{S})$ $\sigma_\mathrm{maj}$ in a unique group of electronic bands centered 2~eV above the VBM in the limit of vanishing magnetic moments ($B$-type ordering). The \ttrue states of PrNiO$_3$ and YNiO$_3$ exhibit different conduction states: for PrNiO$_3$ two separate peaks in the pDOS can be distinguished at 1 and 3~eV above VBM, while in YNiO$_3$ they partially overlap at 2~eV above the same level. This effect is sharply reflected in the magnetization of Ni$_\mathrm{S}$, that decreases substantially from YNiO$_3$ to PrNiO$_3$. We attribute this change in the electronic structure to the larger structural disproportion for the two Ni that is present in the YNiO$_3$ compared to PrNiO$_3$. The \tmeta and \tmetaY states display a strongly reduced band gap. Interestingly in \tmetaY for YNiO$_3$, the system still remains insulating despite the vanishing breathing mode, in contrast with DFT+$U$ where instead the gap closes. 

\subsection{Magnetism-induced electric polarization\label{sec4}} 
\begin{table}[]
\centering
\caption{\label{table2}Electric polarization $\mathbf{P}=\mathbf{P}_\mathrm{e}+\mathbf{P}_\mathrm{I}$ (electronic + ionic) in cartesian axes (the $x$ direction is aligned along $\hat{\textbf{a}}$) in units of $\mu $C/cm$^2$ within the different AFM phases calculated using the DFT+$U$+$V$ method for PrNiO$_3$ and YNiO$_3$, together with the space group of the relaxed structures. The polarization quanta are $(38.3,19.4,27.2)$ $\mu$C/cm$^2$ for PrNiO$_3$ and $(38.8, 21.0,27.6)$ $\mu$C/cm$^2$ for YNiO$_3$. Nonzero components are highlighted in bold.} 
\bgroup
\def\arraystretch{1}
\setlength\tabcolsep{0.09in}
\begin{tabular}{lcccc}
\toprule
\toprule
State&$P_{\mathrm{e},x}$&$P_{\mathrm{e},y}$&$P_{\mathrm{e},z}$&Space group (\#)\\
&$P_{\mathrm{I},x}$&$P_{\mathrm{I},y}$&$P_{\mathrm{I},z}$&\\
\midrule
\multicolumn{5}{c}{PrNiO$_3$}\\
\midrule
$B$&$0.00$&$0.00$&$0.00$&$P2_1/c\,(14)$\\
&$0.00$&$0.00$&$0.00$&\\
\strue&0.00&\textbf{0.21}&$0.00$&$P2_1\,(4)$\\
&0.00&\textbf{0.90}&$0.00$&\\
\ttrue&\textbf{4.25}&0.00&$\textbf{2.74}$&$Pc\,(7)$\\
&$-\textbf{0.86}$&0.00&$-\textbf{2.68}$&\\
\smeta&0.00 & \textbf{1.72}&0.00&$P2_1\,(4)$\\
&0.00 & \textbf{1.58}&0.00&\\
\tmeta&\textbf{11.7}& 0.00&  \textbf{6.14}&$Pc\,(7)$\\
&$-\textbf{1.15}$& 0.00& $-\textbf{4.40}$&\\
\midrule
\multicolumn{5}{c}{YNiO$_3$}\\
\midrule
$B$&$0.00$&$0.00$&$0.00$&$P2_1/c\,(14)$\\
&$0.00$&$0.00$&$0.00$&\\
\strue&0.00&  \textbf{0.07}& 0.00&$P2_1\,(4)$\\
&0.00&  \textbf{0.77}& 0.00&\\
\ttrue&$\textbf{1.27}$&  0.00&  \textbf{0.68}&$Pc\,(7)$\\
&$\textbf{0.13}$&  0.00&  \textbf{0.99}&\\
\smetaY&0.00& $-\textbf{5.57}$& 0.00&$P2_1\,(4)$\\
&0.00& $\textbf{3.18}$& 0.00&\\
\tmetaY&$-\textbf{38.0}$& 0.00&  \textbf{1.02}&$Pc\,(7)$\\
&$-\textbf{0.17}$& 0.00&  \textbf{1.70}&\\
\bottomrule
\bottomrule
\end{tabular}
\egroup
\end{table}
The $P2_1/c$ crystal structure of rare-earth nickelates is centrosymmetric. However, a group theory analysis shows that the magnetic space group of the $S/T$-type magnetic orderings breaks inversion symmetry, thus allowing symmetry-lowering \emph{polar} crystal distortions to occur, resulting in space groups $Pc$ for $T$-type and $P2_1$ for the $S$-type \cite{Giovannetti2009,Ardizzone2021,PerezMato2016}. As described in detail in Ref.~\cite{Ardizzone2021}, the loss of centrosymmetry in the $T$-type ordering is induced by breaking the symmetry of the two-fold screw-axis parallel to $\mathbf{b}$, while for the $S$-type ordering it is due to the loss of a glide plane consisting of a reflection in the plane perpendicular to $\mathbf{b}$ plus a translation along $\mathbf{c}$ of $|\mathbf{c}|/2$.
It has been predicted that the $T$-type ordering would induce an electric polarization $\mathbf{P}$ with nonzero component along $\mathbf{a}$ and $\mathbf{c}$, while the $S$-type ordering an electric polarization predominantly along $\mathbf{b}$~\cite{Giovannetti2009}. The proposed microscopic mechanism consists in a partial charge disproportionation of the oxygens O$_i$ ($i=1,2,3$) forming the octahedra (thus creating two inequivalent O$_i$ and $\mathrm{O}_i'$), which produces a shift from the Ni-site-centered charge disproportionation to a Ni-O-bond-centered charge disproportionation, and therefore allowing a nonzero \textbf{P} to develop. This charge disproportionation of the oxygens is caused by the presence of neighboring nickels with different magnetic moments: either with parallel $\mathrm{Ni}_\sigma$--$\mathrm{O}_i$--$\mathrm{Ni}_\sigma$ or antiparallel $\mathrm{Ni}_\sigma$--$\mathrm{O}_i'$--$\mathrm{Ni}_{-\sigma}$ spins $\sigma=\,\uparrow,\downarrow$. Applying the same argument to the $B$-type ordering, the charge disproportionation of the oxygens can not occur because every O is surrounded by a magnetic and a nonmagnetic Ni atom and this results in a zero electric polarization. 

We analyze the space groups of the relaxed crystal structures and calculate from first-principles the electric polarization of the magnetic orders investigated using DFT+$U$+$V$ method (the $+U$ functional, having vanishing band gaps for $S$ and $T$ types, would yield an ill-defined polarization). As shown in Table \ref{table2}, the relaxed crystal structure of the $B$-type ordering presents a centrosymmetric $P2_1/c$ space group and its electronic ground state does not exhibit any electric polarization, in agreement with our previous considerations. On the contrary, the $S/T$-type AFM orderings display a nonvanishing \textbf{P}, whose magnitude increases with $m(\mathrm{Ni}_\mathrm{S})$. This trend is correlated to the proximity of a (semi)metallic state in the limit of $m(\mathrm{Ni}_\mathrm{S})\rightarrow m(\mathrm{Ni}_\mathrm{L})$. Indeed, an increase of the magnetic moment $m$ is caused by an increase of the difference of the $e_g(\mathrm{Ni})$ occupations for the two spin channels, and thus an enhancement of the energy separation of the associated $e_g(\sigma_\mathrm{min})$ and $e_g(\sigma_\mathrm{maj})$ peaks in the pDOS. Therefore, upon enlarging $m_\mathrm{S}$, the $e_g(\mathrm{Ni}_\mathrm{S})\,\sigma_\mathrm{min}$ states downshift toward the VBM reducing the band gap.

\subsection{Summary}
We studied the structural, electronic and magnetic properties of two members of the rare-earth nickelates series: PrNiO$_3$ and YNiO$_3$. We used DFT functionals augmented with on-site $U$ and inter-site $V$ Hubbard corrections, and we determined the crystal and electronic structures via an iterative self-consistent procedure alternating structural optimizations and evaluations of the Hubbard parameters, until convergence between the two is reached. Calculations were performed without any adjustable parameters, as both Hubbard $U$ and $V$ were calculated from first-principles using density-functional perturbation theory; and we showed that the results change qualitatively upon the inclusion of inter-site Hubbard interactions. We analysed three different antiferromagnetic orderings: \textit{B}, \textit{S} and \textit{T}-types. When employing purely on-site electronic interactions (DFT+$U$), the investigated antiferromagnetic patterns are found to be metastable with respect a ferromagnetic configuration, and some of these (\textit{S}-type and \textit{T}-type orders) yield a semimetallic electronic structure and lose the crystallographic inequivalence (bond disproportionation) of the two Ni sites. Within DFT+$U$+$V$, the results for \textit{B}-type order are very similar to the ones obtained with DFT+\textit{U}. On the other hand, for the $S/T$ antiferromagnetic orderings two different self-consistent states for each magnetic pattern are reached: two metastable states with respect the ferromagnetic solution [\stmeta and \stmetaY for PrNiO$_3$ and YNiO$_3$, respectively] with very similar magnetic moments on the inequivalent Ni sites, reduced breathing mode distortion and small band gaps; and two lowest-energy states with respect to all the magnetic patterns analysed [\sttrue] with substantially different Ni--O bond lengths and magnetic moments for the crystallographic inequivalent nickel atoms, and larger band gaps. In the iterative loop over the calculation of Hubbard parameters and structural optimization, these latter states oscillate between two identical crystal structures with inverted breathing mode amplitudes; we believe that this behaviour might be due to the neglect of the derivative of the Hubbard parameters with respect to the atomic positions and strains \cite{Kulik2011}, and in general for the lack of a functional formulation of the method involving the stationarity with respect to the Hubbard parameters themselves. Finally, without resorting to any empirical parameter, within the DFT+\textit{U}+\textit{V} approach we naturally predict the occurrence of a magnetically-induced electric polarization, thus supporting recent experimental observations pointing to the emergence of a multiferroic phase in this class of materials. %Finally we stress that our results for PrNiO$_3$ and YNiO$_3$, located on opposite sides of the $R$NiO$_3$ phase diagram, are qualitatively similar. This suggests that our conclusions may be extended to the ground state of other members of the rare-earth nickelates series.

\section{methods}
In our first-principles calculations we employ both the rotationally-invariant DFT+$U$ formulation of Dudarev \emph{et al.}~\cite{Dudarev1998}, and the extended DFT+$U$+$V$ one~\cite{Campo2010}. The parameters $U$ and $V$ are calculated from first principles  using DFPT~\cite{Baroni2001,Timrov2018,Timrov2021} as implemented in the \textsc{HP} code~\cite{Timrov2022}. The localized atomic-like set of functions defining the occupation matrices (defined in Ref.~\cite{Cococcioni2005}) employed in our calculations consists of atomic orbitals as found in the pseudopotentials orthogonalized through the L\"{o}wdin algorithm~\cite{Lowdin1950}. For the calculation of the Hubbard contribution to forces and stresses we employ a recently-developed method that uses the solution of the Lyapunov equation to evaluate the derivative of the inverse square root of the overlap matrix~\cite{Timrov2020}. 

All the first-principles calculations are carried out using the Quantum ESPRESSO distribution \cite{Giannozzi2009, Giannozzi2017,Giannozzi2020} . We use a spin-polarized generalized-gradient approximation (GGA) with the PBEsol prescription for the exchange-correlation functional~\cite{Perdew2008}, as well as a PAW pseudopotential (PP) ~\cite{Blochl1994_PAW} for O from the Pslibrary v0.3.1~\cite{DalCorso2014} and ultrasoft (US) PPs ~\cite{Vanderbilt1990} from the GBRV (v1.2) and GBRV (v1.4) libraries~\cite{Garrity2014} for Y and Ni, respectively, as recommended from the SSSP (efficiency) library ~\cite{Prandini2018}. For the Pr we used the US PP  from the Pslibrary1.0.0 (\texttt{Pr.pbesol-spdn-rrkjus\_psl.1.0.0.UPF}) having the $4f$ states frozen in the core. In all our calculations we use 50~Ry as the plane-wave kinetic-energy cutoff for the wavefunctions, 400~Ry kinetic-energy cutoff for the charge density, Gaussian smearing of 0.005~Ry to converge narrow-gap magnetic states and a uniform $\Gamma$-centered 5$\times$8$\times$6 grid of $\mathbf{k}$ points to sample the Brillouin zone. Calculations of the electric polarization are performed using the Berry-phase approach~\cite{King-Smith1993,Umari2002} and using a finer 7$\times$12$\times$8 \textbf{k} mesh. The calculations of the Hubbard parameters are performed using 1$\times$2$\times$2 \textbf{q} points, which corresponds to a supercell of an equivalent size \cite{Timrov2018}. We adopt the iterative procedure detailed in \cite{Cococcioni2019,Timrov2021} to incorporate in the parameters $U$ and $V$ the dependence of both the electronic structure and the crystal environment: this consists in a series of alternating self-consistent structural optimizations and linear-response evaluations of the Hubbard parameters until both the crystal structure and the $U$ and $V$ parameters are converged self-consistently.  

\section{data availability statement}
The data used to produce the results of this paper are available on the Materials Cloud Archive~\cite{MaterialsCloudArchive2022}.

\section{acknowledgements}
We gratefully acknowledge Marisa Medarde for fruitful discussions. This research was supported by the NCCR MARVEL, a National Centre of Competence in Research, funded by the Swiss National Science Foundation (grant number 182892). Computer time was provided by CSCS (Piz Daint) through project No.~s1073.

\section{competing interests}
The authors declare no competing interests.

\section{author contributions}
L.B. performed and analyzed the first-principles calculations,  M.K., I.T. and N.M. helped to interpret and understand the data, while N.M. conceived project. L.B. wrote the initial manuscript and all the authors participated to the final version of it.

\bibliography{bibliography}
\end{document}

\title{Supplementary Information for\\ Hybridization driving distortions and multiferroicity in rare-earth nickelates}

\author{Luca Binci}
\affiliation{Theory and Simulation of Materials (THEOS), and National Centre for Computational Design and Discovery of Novel Materials (MARVEL), École Polytechnique Fédérale de Lausanne, CH-1015 Lausanne, Switzerland}
\author{Michele Kotiuga}
\affiliation{Theory and Simulation of Materials (THEOS), and National Centre for Computational Design and Discovery of Novel Materials (MARVEL), École Polytechnique Fédérale de Lausanne, CH-1015 Lausanne, Switzerland}
\author{Iurii Timrov}
\affiliation{Theory and Simulation of Materials (THEOS), and National Centre for Computational Design and Discovery of Novel Materials (MARVEL), École Polytechnique Fédérale de Lausanne, CH-1015 Lausanne, Switzerland}
\author{Nicola Marzari}
\affiliation{Theory and Simulation of Materials (THEOS), and National Centre for Computational Design and Discovery of Novel Materials (MARVEL), École Polytechnique Fédérale de Lausanne, CH-1015 Lausanne, Switzerland}
\affiliation{Laboratory for Materials Simulations, Paul Scherrer Institut, 5232 Villigen PSI, Switzerland}

%\date{\today}

\maketitle
%\newpage

\section{Spin-resolved projected density of states for Nickel $\bm{t_{2g}}$ states}
\begin{figure*}[h]
\centering
\includegraphics[width=16cm]{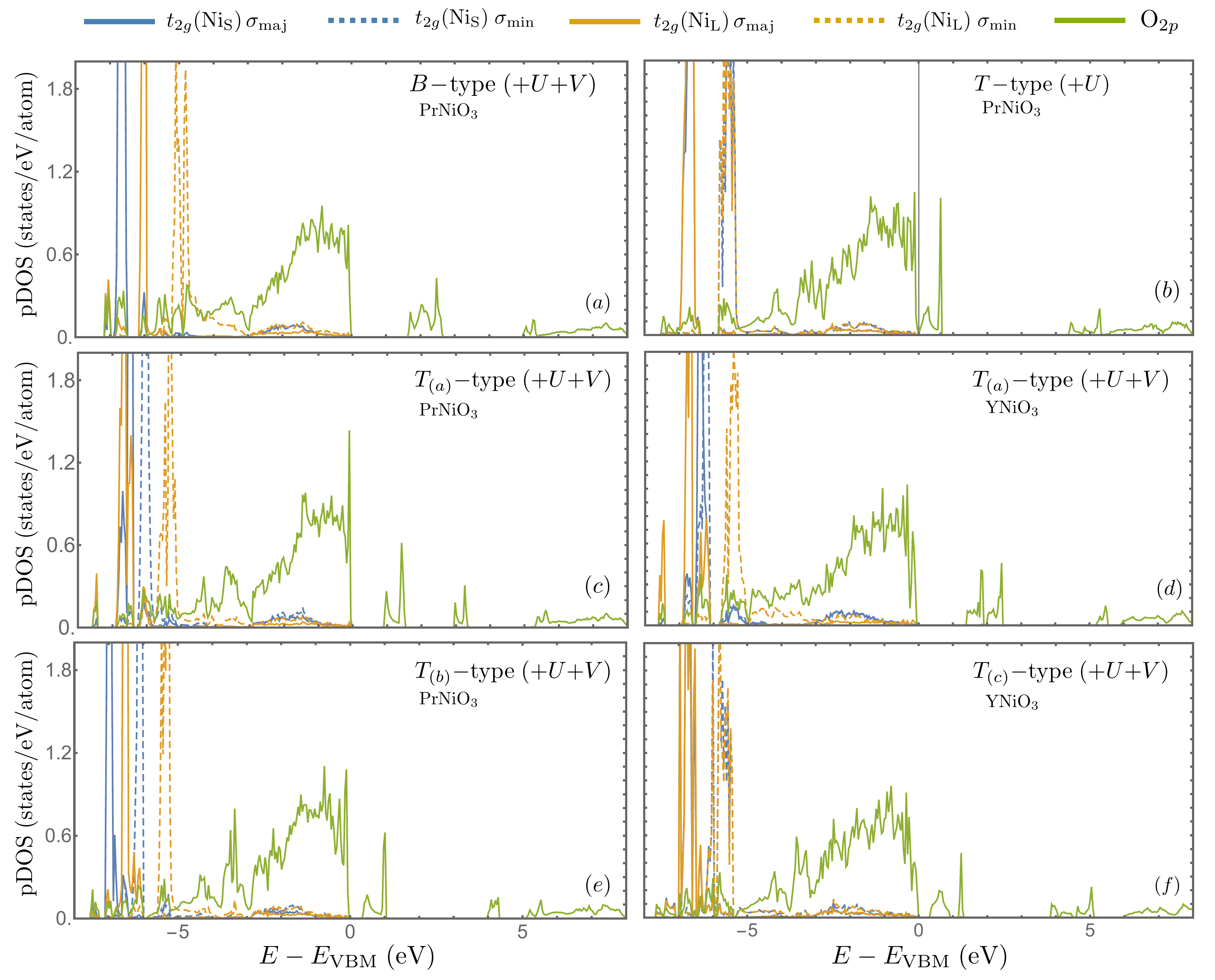}
\caption{pDOS of the relaxed crystal structure for each magnetic ordering. (\textit{a}), (\textit{b}):~\textit{B}-type and $T$-type, respectively, for PrNiO$_3$. (\textit{c}), (\textit{d}):~\ttrue-type and (\textit{e}), (\textit{f}):~\tmeta-type and \tmetaY-type, for both PrNiO$_3$ and YNiO$_3$. The Hubbard functionals employed to produce (\textit{a}), (\textit{c})--(\textit{f}) include both the $U$ and the $V$ interaction parameters, while in (\textit{b}) only the on-site $U$ is present. In the plots the valence Ni--$t_{2g}$ states and the full O--$2p$ projection are shown. We averaged over inequivalent O atoms and over the two Ni($t_{2g}$) states for each spin channel. Given the symmetry between spin-up/spin-down channels for AFM orderings, the notation for majority/minority (maj/min) spin channels is: for a Ni with positive magnetic moments ($m>0$ $\mu_\mathrm{B}$), $\sigma_\mathrm{maj}=\,\uparrow$ and $\sigma_\mathrm{min}=\,\downarrow$; vice versa for Ni with $m<0$ $\mu_\mathrm{B}$. The zero of energy corresponds to the valence band maximum.}
\label{pDOS}
\end{figure*} 

In this section we discuss the pDOS displaying the Ni--$t_{2g}$ states. As shown in Fig.~\ref{pDOS}, the $t_{2g}$ states are located in the range between $-5$ to $-7$~eV with respect to the valence band maximum. These states have a weak hybridization with O-$2p$ states, and since they lie relatively deeply in energy, they are not responsible for the formation of the band gap. This is a consequence of the fact that they are fully occupied, as it can be verified by looking at the eigenvalues $\lambda_{t_{2g},\sigma}$ of the occupation matrices in Table~\ref{table::os}.

\section{Convergence of Hubbard parameters and total energy differences}
\begin{figure*}[h]
\centering
\includegraphics[width=17cm]{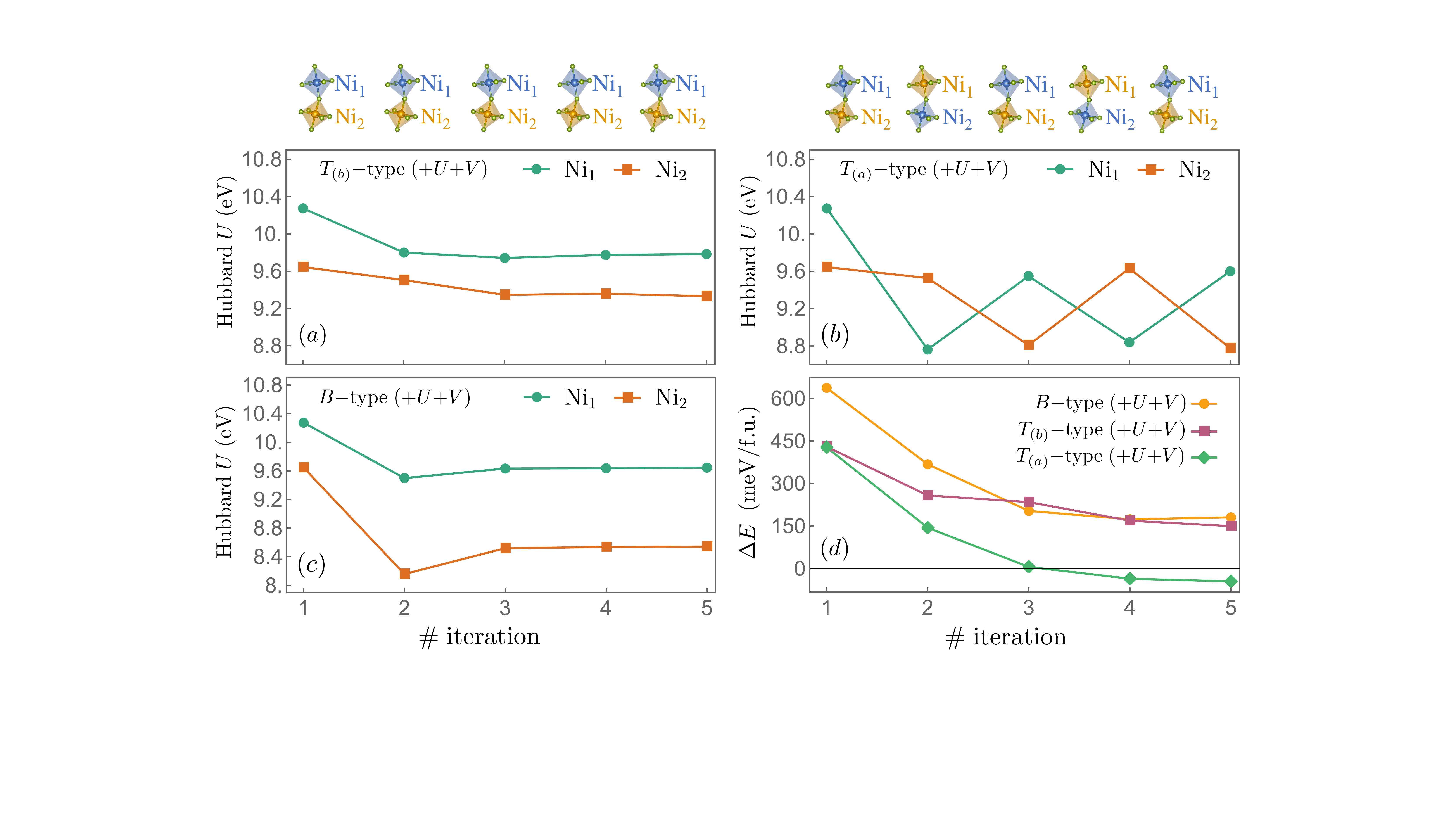}
\caption{Convergence of the Hubbard $U$ parameter (within DFT+$U$+$V$) for the two types of Ni atoms (Ni$_1$ and Ni$_2$) in PrNiO$_3$ for the three types of the AFM magnetic ordering: (\textit{a})~\tmeta, (\textit{b})~\ttrue, and (\textit{c})~$B$. Each iteration of the plot corresponds to a DFPT calculation of $U$ on top of a ground state obtained after a variable-cell structural optimization according to a self-consistent protocol of Ref.~\cite{Timrov2021}. (\textit{d})~Total energy difference $\Delta E = E_\mathrm{AFM} - E_\mathrm{FM}$, where $E_\mathrm{AFM}$ and $E_\mathrm{FM}$ are the total energies of the AFM and FM ground state. On top of panels (\textit{a}) and (\textit{b}), a schematic illustration of the NiO$_6$ volume changes with iterations are shown (blue and yellow octahedra correspond to larger and smaller volumes, respectively).}
\label{fig:conv}
\end{figure*} 
In this section we show the convergence of the Hubbard parameters and total energy differences for the different types of the magnetic ordering discussed in the main text for the PrNiO$_3$ material. The convergence for YNiO$_3$ follows similar trends. The Hubbard parameters are computed using the self-consistent protocol that is described in detail in Refs.~\cite{Cococcioni2019,Timrov2021}. This workflow consists in the alternate evaluation of Hubbard parameters and structural optimization until convergence is achieved within a certain accuracy. 

Fig.~\ref{fig:conv}~(\textit{a})--(\textit{c}) show the convergence of Hubbard $U$ during the iterations of the self-consistent cycle for PrNiO$_3$. The \tmeta-type (+$U$+$V$) and the $T$-type (+$U$) orderings display converged values of $U$ and Ni--O octahedral volumes. On the contrary, the \ttrue-type exhibit oscillations of the Hubbard $U$ values, that is associated to a similar behaviour of the octahedral volumes of the crystallographically inequivalent Ni$_1$/Ni$_2$, as schematically shown in the cartoon above panel (\textit{b}) in the same figure. Similar trends occur also for the magnetic moments in the Ni atoms and on the Hubbard $V$ parameters. A possible cause for this swapping could be due to the neglect of the derivative of the Hubbard parameters with respect to the atomic displacements \cite{Kulik2011}. Indeed, focusing on the leading on-site term, such contribution to the force would be $\sum_{I}\frac{dU^I}{d\mathbf{R}}\mathrm{tr}\,(\mathbf{n}^I(\mathbf{1}-\mathbf{n}^I))$. Since the trace gives a positive contribution while $U^I$ increases (decreases) upon contraction (expansion) of the Ni-O octahedra, the net result is that this term opposes to the mechanism of swapping during the structural optimization. A quantification of this component to the Hubbard force, although possibly significant, is however beyond the scope of this study. We remark that, although the breathing mode and the Hubbard parameters swap, the total energy of the \ttrue state smoothly converges and -- after some initial iterations -- it stabilizes as the lowest energy magnetic state, as displayed in panel (\textit{d}) of Figure~\ref{fig:conv}. In addition, together with the Hubbard parameters, also the magnetic moments and the Ni--O volumes octahedra associated to the Ni$^\mathrm{S}_i$ and Ni$^\mathrm{L}_i$ ($i=1,2$) are numerically always the same. We therefore conclude that although the apparently oscillating behaviour, the final result is unambiguous.

\section{Band gaps}
In Table~\ref{tab:gaps} we show the band gap values for PrNiO$_3$ and YNiO$_3$ computed using DFT+$U$ and DFT+$U$+$V$ at the self-consistent electronic and crystal structures. Experimentally, both materials are known to be insulating in the investigated phase \cite{Medarde1997}. The insulating character is obtained for all types of the magnetic ordering when the DFT+$U$+$V$ functional is used, while only the AFM-$B$ type is insulating within DFT+$U$. 
%Therefore, the intersite Hubbard $V$ interactions are pivotal for the correct prediction of the insulating nature of the $T_\mathrm{true}$ and $T_\mathrm{meta}$ AFM magnetic orderings.For the $B$ type we observe that DFT+$U$+$V$ gives slightly larger (by $0.2-0.3$~eV) band gaps that at the DFT+$U$ level of theory. Within DFT+$U$+$V$, we find that YNiO$_3$ has a larger band gap than PrNiO$_3$ when considering the $S_\mathrm{true}$ and $S_\mathrm{meta}$ types, while the trend is the opposite for the $T_\mathrm{true}$ type and gaps are almost equal for the $T_\mathrm{meta}$ type. In order to shed more light on such comparisons of the band gaps for these two materials and different magnetic orderings, new high-resolution experiments are needed and thus are called for.  

\begin{table*}[h]
\centering
\caption{Values of the band gaps for different magnetic orderings calculated with the two different methods DFT+$U$ and DFT+$U$+$V$ for $R$NiO$_3$ ($R=$ Pr and Y).}
\bgroup
\def\arraystretch{1}
\setlength\tabcolsep{0.18in}
\begin{tabular}{lccccccccc}
\toprule
\toprule
&\multicolumn{3}{c}{DFT+$U$}&&\multicolumn{5}{c}{DFT+$U$+$V$}\\
\midrule
%\cline{1-3}\cline{5-9}
PrNiO$_3$&$S$&$T$&$B$&&\strue&\ttrue&\smeta&\tmeta&$B$\\
$E_\mathrm{gap}$ (eV)&0.00&0.00&1.37&&0.92&0.89&0.36&0.35&1.64\\
\midrule
YNiO$_3$&$S$&$T$&$B$&&\strue&\ttrue&\smetaY&\tmetaY&$B$\\
$E_\mathrm{gap}$ (eV)&0.00&0.00&1.41&&1.42&1.33&0.27&0.34&1.65\\
\bottomrule
\bottomrule
\end{tabular}
\egroup
\label{tab:gaps}
\end{table*}

\section{L\"owdin occupation numbers and oxidation states}

In table~\ref{table::os} we show the eigenvalues $\lambda^I_{m,\sigma}$ of the occupation matrices $n^{I\sigma}_{mm'}=\sum_i\,f_i\times\langle\psi_{i\sigma}|\phi_{m'}^I\rangle\langle\phi_m^I|\psi_{i\sigma}\rangle$ \cite{Cococcioni2005}. We also provide the values of the oxidation states following the prescription of Ref.~\cite{Sit2011}. This is based on considerations about occupations of bonding/antibonding states when the atomic orbitals of a transition-metal (TM) and a ligand mix; according to~\cite{Sit2011}, only full $d$-occupancies ($\lambda_{m,\sigma}^I\gtrsim0.9$) of the TM orbitals provide evidence that an electron is effectively bound to the ion, whereas considering only the total occupation $n^I=\sum_{m,\sigma}n^{I\sigma}_{mm}$ could lead to misleading conclusions. 

\begin{table*}[h]
\centering
\label{table::os}
\caption{Eigenvalues ($\lambda_{i,\sigma}$, $i=1,\dots,5$) of the occupation matrices for the majority/minority spin channels $(\sigma_\mathrm{maj}/\sigma_\mathrm{min})$ of the Ni--$3d$ manifold, both for the long and short bonds (L/S) Ni atoms, for different magnetic orderings. All the results refers to DFT+$U$+$V$ calculations, except for $T^{(+U)}$ that is obtained with DFT+$U$. In the table, $n=\sum_{i,\sigma}\lambda_{i,\sigma}$ and OS refers  to the oxidation state.} 
\bgroup
\def\arraystretch{1.0}
\setlength\tabcolsep{0.07in}
\begin{tabular}{lcccccccccccccccc}
\toprule
\toprule
&&\multicolumn{5}{c}{$\lambda_{i,{\sigma_\mathrm{maj}}}$}&&\multicolumn{5}{c}{$\lambda_{i,{\sigma_\mathrm{min}}}$}\\
%\cline{3-7}\cline{9-13}
State&Ni&&$e_g$ &$e_g$ &$t_{2g}$ &$t_{2g}$ &$t_{2g}$ &&$e_g$ &$e_g$ &$t_{2g}$ &$t_{2g}$ &$t_{2g}$&&$n$&OS \\
\midrule
%$B$&S&0.54&  0.55&  \textbf{0.99}&  1.00&  1.00&&0.54& 0.55&  0.99&  1.00&  1.00&8.16&0.00&Ni(IV)\\
%&L&0.98&  0.98&  0.99&  1.00&  1.00&&0.13&  0.14&  0.99&  0.99&  0.99&8.18&1.71&Ni(II)\\
%$T^{(+U)}$&S&0.91&  0.95&  0.99&  1.00&  1.00&&0.14&  0.16&  0.99&  0.99&  0.99&8.12&1.57&Ni(II)\\
%&L&0.92&  0.95&  0.99&  1.00&  1.00&&0.14&  0.16&  0.99&  0.99&  0.99&8.13&1.59&Ni(II)\\
%\tmeta&S&0.83&  0.86&  1.00&  1.00&  1.00&&0.24&  0.25&  0.99&  0.99&  0.99&8.13&1.20&Ni(IV)\\
%&L&0.96&  0.97&  0.99&  1.00&  1.00&&0.14&  0.15&  0.99&  0.99&  0.99&8.16&1.65&Ni(2)\\
%\ttrue&S&0.72&  0.73&  0.99&  1.00&  1.00&&0.36&  0.37&  0.99&  0.99&  1.00&8.14&0.72&Ni(IV)\\
%&L&0.98&  0.98&  0.99&  1.00&  1.00&&0.13&  0.13&  0.99&  0.99&  0.99&8.16&1.72&Ni(II)\\
%\midrule
%\multicolumn{16}{c}{YNiO$_3$}\\
%\midrule
%\tmeta&S&0.80&  0.98&  0.99&  1.00&  1.00&&0.14&  0.26&  0.99&  0.99&  0.99&8.14&1.38&Ni(III)\\
%&L&0.80&  0.98&  0.99&  1.00&  1.00&&0.14&  0.26&  0.99&  0.99&  0.99&8.14&1.38&Ni(III)\\
%\ttrue&S&0.58&  0.60&  0.99&  1.00&  1.00&&0.49&  0.51&  0.99&  1.00&  1.00&8.15&0.19&Ni(IV)\\
%&L&0.98&  0.99&  0.99&  1.00&  1.00&&0.12&  0.13&  0.99&  0.99&  0.99&8.17&1.74&Ni(II)\\
\multicolumn{17}{c}{PrNiO$_3$}\\
\midrule
$B$&S&&0.54&  0.55&  0.99&  1.00&  1.00&&0.54& 0.55&  0.99&  1.00&  1.00&&8.16&Ni(IV)\\
&L&&0.98&  0.98&  0.99&  1.00&  1.00&&0.13&  0.14&  0.99&  0.99&  0.99&&8.18&Ni(II)\\
$T^{(+U)}$&S&&0.91&  0.95&  0.99&  1.00&  1.00&&0.14&  0.16&  0.99&  0.99&  0.99&&8.12&Ni(II)\\
&L&&0.92&  0.95&  0.99&  1.00&  1.00&&0.14&  0.16&  0.99&  0.99&  0.99&&8.13&Ni(II)\\
\tmeta&S&&0.83&  0.86&  1.00&  1.00&  1.00&&0.24&  0.25&  0.99&  0.99&  0.99&&8.13&Ni(IV)\\
&L&&0.96&  0.97&  0.99&  1.00&  1.00&&0.14&  0.15&  0.99& 0.99&  0.99&&8.16&Ni(II)\\
\ttrue&S&&0.72&  0.73&  0.99&  1.00&  1.00&&0.36&  0.37&  0.99&  0.99&  1.00&&8.14&Ni(IV)\\
&L&&0.98&  0.98&  0.99&  1.00&  1.00&&0.13&  0.13&  0.99&  0.99&  0.99&&8.16&Ni(II)\\
\midrule
\multicolumn{17}{c}{YNiO$_3$}\\
\midrule
\tmetaY&S&&0.80&  0.98&  0.99&  1.00&  1.00&&0.14&  0.26&  0.99&  0.99&  0.99&&8.14&Ni(III)\\
&L&&0.80&  0.98&  0.99& 1.00&  1.00&&0.14&  0.26& 0.99&  0.99&  0.99&&8.14&Ni(III)\\
\ttrue&S&&0.58&  0.60&  0.99& 1.00&  1.00&&0.49&  0.51&  0.99&  1.00&  1.00&&8.15&Ni(IV)\\
&L&&0.98&  0.99&  0.99&  1.00&  1.00&&0.12&  0.13&  0.99&  0.99&  0.99&&8.17&Ni(II)\\
\bottomrule
\bottomrule
\end{tabular}
\egroup
\end{table*}
\bibliography{bibliography}